\documentclass[aps,prl,twocolumn,groupedaddress,amsmath,showpacs]{revtex4}
\usepackage{graphicx}
\usepackage{subfigure}

\begin{document}


\title{A steady state network model with a 1/k scale-free degree distribution}


\author{Simon Laird}
\affiliation{Department of Mathematics, Imperial College London, 180 Queen's Gate, London SW7 2AZ, U.K.}
\author{Henrik Jeldtoft Jensen}
\email[Author to whom correspondence should be addressed:\\]{h.jensen@imperial.ac.uk}
\homepage{http://www.ma.ic.ac.uk/~hjjens/}
\affiliation{Department of Mathematics, Imperial College London, 180 Queen's Gate, London SW7 2AZ, U.K.}


\date{\today}

\begin{abstract}
Using a steady state process of node duplication and deletion we produce networks with $1/k$ scale-free degree distributions in the limit of vanishing connectance. This occurs even though there is no growth involved and inherent preferential attachment is counterbalanced by preferential {\emph{detachment}}. The mean field evolution is considered and the $1/k$ law is verified under certain approximations. An ansatz for the degree distribution is proposed on the basis of symmetry considerations and is shown to coincide well with the simulation data. Distributional forms other than power law are also shown to arise when the condition of perfect duplication is relaxed.   
\end{abstract}


\maketitle


The frequent appearance of power-law degree distributions in real system networks is an intriguing phenomenon. It has been suggested that the scale-free nature of such inter-connected systems offers a level of robustness and stability to the dynamics \cite{doro02:evol,albe02:stat} so is favoured in a selective sense. Whilst this may be true the presence of these distributions may actually be attributed to the manner in which the networks undergo construction.

The mechanism of preferential attachment \cite{bara99:emer} has been used successfully to model the development of many non-biological networks \cite{albe02:stat,chun03:dupl}, reproducing power laws with the correct range of exponents. For biological systems, the process of node duplication is more appropriate and models of this type have again been efficacious \cite{bhan02:dupl,kim02:infi,vazq03:mode,past03:evol,coli05:char,ispo05:dupl}. These two mechanisms have a level of equivalence though \cite{coli05:char} and a common theme amongst their application is the reliance upon node growth in the dynamics.

Approaches using fixed node numbers or node deletions have been proposed that use a variety of mechanisms to achieve scale-free and other distributions \cite{doro00:scal,albe00:topo,epps02:stea,sars04:scal,deng04:grow,sala05:evol}. In this letter we propose a simple model of duplication and deletion where the node number remains constant throughout the network evolution. The duplication process can be incurred perfectly or with a relaxation in the fidelity in order to represent partially correlated duplications.

With perfect duplication our model produces a $1/k$ power law degree distribution in the limit of vanishing connectance. An ansatz for the functional form of the degree distribution is proposed based upon symmetry considerations and is fitted well by the simulation data. With certain approximations we provide a solution to the mean field degree evolution equation which supports the presence of an exponent, $\gamma{=}1$. When we relax the condition of perfect duplication we see distributions better described by the exponential form. Ecological field data often displays such distributions \cite{dunn02:netw,jord03:inva} and a previous model of species interaction networks has presented them also \cite{lair05:tang}. It could be argued that ecological systems develop via processes of extinction and correlated speciation with steady state species numbers so the appearance of similar degree distributions here is of relevance.

{\it The model - } We consider a network of $N$ nodes where edges are undirected and self-connection is excluded. The initialised edge set is transient under the proposed dynamics so can be taken from any arbitrary distribution.
 
The edge evolution progresses via non-local updates but is nevertheless very simple. At a timestep, we randomly select with uniform probability a node and all its associated edges to be deleted from the network. A second {\emph{parent}} node is then randomly selected from the remaining $N{-}1$ nodes and is duplicated in the form of a {\emph{daughter} node. All nodes connected to the parent are now given connections to the daughter with probability $P_{e}$. All nodes not connected to the parent are given connections to the daughter with probability, $P_{n}$. The determination of an edge between the daughter and the parent is made with probability, $P_{p}$.
\begin{figure}[b]
\centering
\includegraphics[]{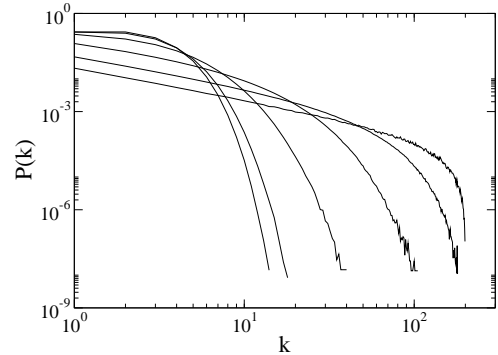}
\caption{\small{Degree distributions for $N=200$, ${C}_{0}=0.01$ systems produced using the imperfect duplication process. From short to long tail we have $P_{e}{=}\{0.01,0.25,0.75,0.95,0.99,0.999\}$.}}\label{fig.id1}
\end{figure}
As will be shown, at mean field level the system admits a single attracting fixed point in the connectance, $\bar{C}{\to}C_{0}$, which is determined by the duplication probabilities (Eq.\ref{eq.theta}). By using this relation we select the probabilities, $P_{e},P_{n}$, dependently such that $P_{p}{=}C_{0}$, so allowing $P_{e}{\in}[{C_{0}},1]$ to be used as a control parameter for the fidelity. As a result, the dynamics range from fully random, $P_{e}{=}P_{n}{=}P_{p}{=}C_{0}$, generating binomial networks, to fully correlated, $P_{e}{=}1, P_{n}{=}0, P_{p}{=}C_{0}$, where daughter and parent are indistinguishable.

The ensemble averaged degree distributions for various fidelities are shown in Fig.(\ref{fig.id1}). For the uncorrelated, random duplication process, $P_{e}{=}P_{n}{=}C_{0}$ we see binomially distributed networks as expected. As we increase the fidelity, the distributions exhibit longer tails and conform more to exponential than they do binomial. This progression continues until we have a process of perfect duplication at which point the distributions become power-law-like.

Shown in Fig.(\ref{fig.pd1}), are the ensemble averaged degree distributions for perfect duplication networks with ${C}_{0}=0.01$ over increasing system sizes. For finite but small values of the connectance the distributions achieve forms that conform closely to a power law. As we reduce ${C}_{0}$ towards zero, the functional forms become strongly power-law-like with exponents, $\gamma{\simeq}1$. The networks achieved here are extremely sparse and the majority of the support is held in the $P(k{=}0)$ degree. So the form we see here represents (near) scale-free fluctuations above the $k{=}0$ vacuum. In the limit that ${C}_{0}{\rightarrow}{0}$ a finite sized system asymptotically achieves the absorbing state of zero connectance. As we take the thermodynamic limit in the network size, though, we have a system where $P(k{=}0)$ tends to unity and the remaining distribution achieves power-law form but with the support for $P(k{>}0)$ tending to zero. The simulations suggest that in these limits we have a network of zero connectance with $1{/}k$ scale-free fluctuations in the degree. It is also interesting to observe a mirror symmetry between distributions of complementary connectance (${{C}_{0}}'{=}1{-}{{C}_{0}}$), in Fig.(\ref{fig.pd2}). The distribution for ${C}_{0}{=}0.49$ is also shown, demonstrating their gradual convergence to the same uniform profile at ${C}_{0}{=}0.5$.   
\begin{figure}
\centering
\includegraphics[]{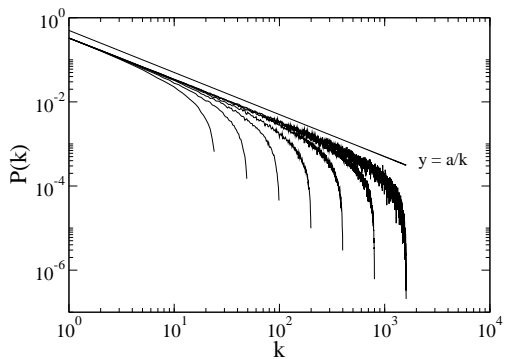}
\caption{\small{Perfect duplication degree distributions for networks with $N{=}\{25,50,100,200,400,800,1600\}$, ${C}_{0}{=}0.01$. Each has been normalised with the exclusion of the ${k}{=}{0}$ support and rescaled to overlap for visual purposes.}}\label{fig.pd1}
\end{figure}

\begin{figure}[b]
\centering
\subfigure{
\includegraphics[]{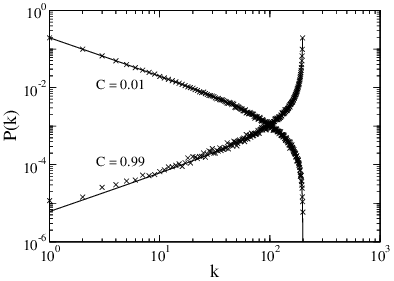}}
\subfigure{
\includegraphics[]{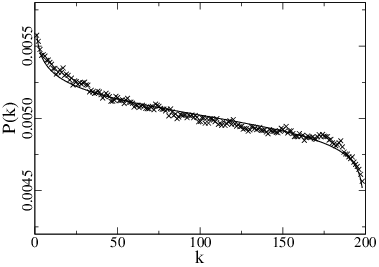}}
\caption{\small{Perfect duplication degree distributions for $N{=}200$, ${C}_{0}{=}0.01,0.99$ (top); ${C}_{0}{=}0.49$ (bottom). Curve fits are applied using Eq.(\ref{eq.degdist}).}} \label{fig.pd2}
\end{figure}

{\it Symmetries - } We propose a functional form for the degree distributions acquired with the perfect duplication process, based upon symmetry requirements from both theoretical and empirical information. Note that the probability of duplicating an edge, $P_{e}{=}1$, whilst the probability of duplicating a lack of edge is, $1{-}P_{n}{=}1$. Additionally, the parent-daughter edge is created with probability, $P_{p}$, and complementary to this a non-edge is, $1{-}P_{p}$. There are no other aspects to consider in this process. Therefore, if we shift our frame of reference from edge-centric to non-edge-centric we may state that the degree distribution has the following symmetry,
\begin{equation}
P(k';P_{p},N)=P(1{-}k';1{-}P_{p},N),
\label{eq.sym1}
\end{equation}
where we have used a scaled variable for the degree, $k'{=}\frac{k}{N{-}1}$. This duality relation is quite general with multiple classes of solutions but we can constrain the possibilities by using the empirically observed mirror symmetry. When comparing distributions of complementary connectance ${{C}_{0}}'{=}1{-}{{C}_{0}}$ (Fig.\ref{fig.pd2}), they appear to be mirror symmetric versions of one another on the log-log scale, a feature that can be represented by the relation,
\begin{equation}
log P(k';P_{p},N) = {-} log P(k';1{-}P_{p},N){+}A(P_{p},N),
\label{eq.sym2a}
\end{equation}
where $A(P_{p},N)$ is a constant for a given $P_{p}$ and $N$. If we use the previous symmetry of Eq.(\ref{eq.sym1}), this may be reformulated as,
\begin{equation}
P(k';P_{p},N)P(1{-}k';P_{p},N)=A(P_{p},N).
\label{eq.sym2b}
\end{equation}
So with these two symmetries in mind we now propose an ansatz for the degree distribution of the form,
\begin{equation}
P(k';P_{p},N)=
\begin{cases}
P(0;P_{p},N), & \text{$k'{=}0$;}\\
G(P_{p},N)[\frac{k'}{1{-}k'}]^{{-}(1{-}2P_{p})}, & \text{$0{<}k'{<}1$;}\\
P(1;P_{p},N), & \text{$k'{=}1$.}
\end{cases}\label{eq.degdist}
\end{equation}
This is consistent with both the theoretical and observed symmetries and agrees strongly with the distributions acquired from the simulations (Fig.\ref{fig.pd2}). For the uniform degree distribution of ${C}_{0}{=}0.5$ our functional form also fits well. The exponent of the function tends to zero at this point leaving $P(k';0.5,N){=}constant$ as required.

{\it Mean field theory - }To demonstrate the existence of an attracting fixed point in the connectance we now consider 
the mean field dynamics of the number of edges, $E_{t}$,
\begin{eqnarray}
E_{t{+}2} &=& E_{t}{-}\bar{k}_{t}{+}{P_{e}}\bar{k}_{t{+}1}{+}{P_n}(N{-}2{-}\bar{k}_{t{+}1}){+}P_{p},\nonumber\\
\bar{k}_{t} &=& \frac{2E_{t}}{N},\,\,\,
\bar{k}_{t{+}1} = \frac{2}{N{-}1}(E_{t}{-}\bar{k}_{t}).\label{eq.mf_evo1}
\end{eqnarray}
The second term accounts for the loss of edges when a node with mean degree, $\bar{k}_{t}$ is deleted in stage one. The third and fourth terms then represent the increase in edges incurred at stage two when one of the remaining $N{-}1$ nodes is duplicated. Imperfect duplication is made possible at this point by the inclusion of the probabilities, $P_{e}$ and $P_{n}$. The fifth term represents the contribution from the daughter-parent connection. 

As we only observe the system after every two steps we may rescale the time variable increment, $(t{+}t'){\to}(t{+}\frac{t'}{2})$. Thus Eq.(\ref{eq.mf_evo1}) may be expressed as a simple recursion relation which admits a single attracting fixed point that can be expressed in terms of the network connectance,
\begin{equation}
{\bar{C}_{0}}=\frac{P_{n}(N{-}2){+}P_{p}}{N{-}1{-}(P_{e}{-}P_{n})(N{-}2)}.
\label{eq.theta}
\end{equation}

We now consider the evolution equation for the average number of nodes of degree k, $n_{k}$, constrained by the condition, $\sum_{k}n_{k}{=}N$,
\begin{equation}
n_{k}(t{+}1)=n_{k}(t) + \Gamma_R(N,k,t) + \Gamma_{Du}(N{-}1,k,t). 
\label{master_D}
\end{equation}
Here the probabilities, $\Gamma_R(N,k,t)$ and $\Gamma_{Du}(N{-}1,k,t)$ concern the effects of removal (R) and duplication (Du) respectively. They consist of the following contributions (we suppress for convenience the timestep variable $t$ and the size of the network, $N$ or $N{-}1$):
\begin{eqnarray}
\Gamma_R(k)    &=& -\Gamma^r_R(k) +\Gamma^a_R(k{+}1)-\Gamma^a_R(k).\\
\Gamma_{Du}(k) &=& \Gamma^p_{Du}(k{-}1)-\Gamma^p_{Du}(k)+\Gamma^d_{Du}(k)\nonumber\\
               & &+\Gamma^a_{Du}(k{-}1)-\Gamma^a_{Du}(k).
\end{eqnarray}

The separate terms represent the effects of the process on the distinct node types, indexed as: removed (r), adjacent (a), parent (p), daughter (d).

The direct effect on $n_{k}$ of removing a node of degree, $k$ is to decrease its support by one. The probability of selecting a node of degree $k$ is $\frac{n_{k}}{N}$, and therefore,
\begin{equation}
\Gamma^r_R(k)=\frac{n_{k}}{N}.
\end{equation}
Upon removal, the degree of the nodes connected to the removed node will decrease by one. For this we need the {\em Edge} probability, $P_{Ed}(k_1,k_2,q)$, the probability that a node of degree $k_{1}$ is connected to $q$ nodes of degree $k_{2}$. This will be estimated below (Eq.\ref{eq.pedge}). Here we write,
\begin{equation}
\Gamma_R^a(k)=\sum_{k_r=1}^{N{-}1}\frac{n_{k_r}}{N}\sum_{q=1}^{k_r}qP_{Ed}(k_r,k,q).
\end{equation}
The first sum is over the degree of the removed node, the second sum is over the number of nodes of degree $k=0,1,...,N{-}1$ the removed node is connected to.

When a daughter is introduced it receives an edge to the parent with probability $P_p$. Thus the parent increases its degree by one. A node of degree $k$ is selected for duplication with probability $n_{k}/(N{-}1)$,
\begin{equation}
\Gamma_{Du}^p(k)=P_p\frac{n_{k}}{N{-}1}.
\end{equation}
The daughter can add support to $n_{k}$ by an amount given by the probability of finishing with $k$ edges,
\begin{equation}
\Gamma_{Du}^d(k) =P_p\Lambda(k{-}1)+(1{-}P_p)\Lambda(k).
\end{equation}
where,
\begin{gather}
\begin{split}
\Lambda(k) &= {\rm Prob}\{\rm daughter \; receives \; 
{\it k} \; edges \; to 
\; nodes \\
& {\quad} {\rm \; different \; from \; the \; parent} \}
\end{split}\\
=\sum_{k_p=0}^{N-2}\sum_{k_1=0}^{{\rm m}[k_p,k]}
\sum_{k_2=0}^{{\rm m}[N{-}2{-}k_p,k]}
\frac{n_{k_p}}{N{-}1}\delta(k_1{+}k_2{-}k)\Omega(k_1,k_2,k_p),\nonumber
\end{gather}
where {\emph m} in the limits represents, minimum. This expression corresponds to when the daughter inherits $k_1$ edges associated with nodes connected to the parent with each attachment occurring with probability, $P_{e}$. The remaining $k_2{=}k{-}k_1$ edges are established with probability, $P_n$ and are associated with nodes not connected to the parent. The factor $\Omega(k_1,k_2,k_p)$ is the probability that $k_1$ edges derive from the $k_p$ nodes connected to the parent and $k_2$ from those not connected to the parent,
\begin{equation}
\begin{split}
\Omega(k_1,k_2,k_p) &=\left(\begin{array}{c} k_p\\ k_1\end{array}\right)
P_e^{k_1}(1{-}P_e)^{k_p{-}k_1}\\
& {\quad} \left(\begin{array}{c} N{-}2{-}k_p\\ {k_2}\end{array}\right)    
P_n^{k_2}(1{-}P_n)^{N{-}2{-}k_p{-}k_2}.
\end{split}
\end{equation}
The duplication process also effects potential adjacent nodes connected or unconnected to the parent. A parent of degree $k_p$ is connected to $q$ nodes of degree $k$ with probability $P_{Ed}(k_p,k,q)$. These $q$ nodes will as a result of the duplication with probability $P_e$ become degree $k{+}1$ nodes. There are $N{-}2{-}k_p$ nodes the parent has no link to. Among these nodes $q$ will receive a new edge to the daughter and become degree $k{+}1$ nodes with probability $P_{n}P_{Ed}(N{-}2{-}k_p,k,q)$. The total contribution to the transition probability is,
\begin{eqnarray}
\Gamma_{Du}^a(k)&=&
\sum_{q=0}^{N{-}2}\sum_{k_p=0}^{N{-}2}\sum_{\kappa_1=0}^{k_p}
\sum_{\kappa_2=0}^{N{-}2{-}k_p}\sum_{q_1=0}^{\kappa_1}\sum_{q_2=0}^{\kappa_2}
\delta(q_1{+}q_2{-}q)\nonumber \\
&&q\frac{n_{k_p}}{N{-}1}P_{Ed}(k_p,k,\kappa_1)
\left(\begin{array}{c} \kappa_1\\ q_1\end{array}\right)
P_e^{q_1}(1{-}P_e)^{\kappa_1{-}q_1}\nonumber \\
&&P_{Ed}(N{-}2{-}k_p,k,\kappa_2)
\left(\begin{array}{c} \kappa_2\\ 
q_2\end{array}\right)
P_n^{q_2}(1{-}P_n)^{\kappa_2{-}q_2}.\nonumber \\
\label{eq.final}
\end{eqnarray}
We estimate $P_{Ed}(k_1,k_2,q)$ by using the following urn-model consideration. We have as many particles in the urn as there are edges, i.e, $M{=}\sum_k kn_k$. We imagine that the particles are of two types, $A$ and $B$. The former corresponds to $|A|{=}k_2(n_{k_2}{-}\delta_{k_1,k_2})$; 
the edges connecting to nodes of degree $k_2$. The latter to $|B|{=}M{-}|A|$; the edges leading to nodes with degree different from $k_2$. We pick $k_1$ particles from the urn. These are the edges which will be connected to the single node of degree $k_1$ under consideration. The probability that among these $k_1$ particles, $q$ are of type $A$ and $k_1{-}q$ are of type $B$ is,
\begin{equation}
P_{Ed}(k_1,k_2,q){=} \left(\begin{array}{c} k_1\\ q\end{array}\right)
\big(\frac{|A|}{M}\big)^q
\big(1{-}\frac{|A|}{M}\big)^{k_1{-}q}.\label{eq.pedge}
\end{equation}
It is straightforward to check that these equations support the normalised stationary solution $P(k)\,{\propto}\,{1/k}$ in the double limit $P_n{\rightarrow}{0}$ and $P_d{\rightarrow}{1}$, when only leading contributions in the sum over $P_{Ed}$ from Eq.(\ref{eq.final}) are included. This implies the relevant correlations make $P_{Ed}(k_1,k_2,q)$ decay rapidly to zero with increasing $k_1$ and $q$. Measured degree-degree correlations in the numerics support this.

{\it Conclusion - }We have shown that a power-law degree distribution may be achieved without the usual requirement of node growth. The ansatz for the degree distribution suggests that the associated exponent is equal to unity, and the solution to the degree evolution equation in the required limits supports this. Both the processes of preferential attachment and duplication can give a range of exponents but achieving an exponent this low is unusual. Biological systems and in particular ecological systems often exhibit low range exponents when power laws occur, $\gamma\simeq(1,2)$ \cite{chun03:dupl}, so it is interesting to see this reproduced in our model. Ecological systems often present degree distributional forms other than the power-law and our model shows that partially correlated duplication can lead to distributions better described as exponential. The appearance of non-binomial distributions in ecology may just be an inherent consequence of the system dynamics. 

The duplication process in growth models naturally incorporates preferential attachment which allows a strong comparison between the two approaches. But for our model the equivalence between the two mechanisms becomes blurred. By utilising deletion as well as duplication, higher degree nodes again have greater probabilities of increasing their degree during duplication events. But conversely they have greater probabilities of reducing their degree during the deletion events ie. preferential {\emph {detachment}}. The 'more begets more' adage seems inappropriate here as more also begets less.

\begin{acknowledgments}
We wish to thank Andrew Parry, Kim Christensen, Gunnar Pruessner and the Imperial College Complexity Group for valuable insights. A debt of thanks goes to Andy Thomas for computational support. Simon Laird thanks the EPSRC for PhD funding of this work.
\end{acknowledgments}

\bibliography{paper_PRL}

\end{document}